\documentclass[twocolumn, superscriptaddress, nofootinbib, showpacs,
amsmath, amssymb]{revtex4}

\begin{document}

\title{ Free particle wavefunction in light of the minimum-length deformed quantum mechanics and some of its phenomenological implications }

\author{ Micheal S. Berger }
\email{berger@indiana.edu}\affiliation{Physics Department, Indiana University, Bloomington, IN 47405, USA}
\author{ Michael~Maziashvili}
\email{maziashvili@gmail.com} \affiliation{ Andronikashvili
Institute of Physics, 6 Tamarashvili St., Tbilisi 0177, Georgia}  \affiliation{ Center for Elementary Particle Physics, ITP, Ilia State University, \\ 3\,$-$\,5 Cholokashvili Ave., Tbilisi 0162, Georgia }

\begin{abstract}

At a fundamental level the notion of particle (quantum) comes from quantum field theory. From this point of view we estimate corrections to the free particle wave function due to minimum-length deformed quantum mechanics to the first order in the deformation parameter. Namely, in the matrix element $\langle 0 |\varPhi(t,\,\mathbf{x}) |\mathbf{p} \rangle$ that in the standard case sets the free particle wave function $\propto \exp\left(i\left[\mathbf{p}\mathbf{x} - \varepsilon(\mathbf{p}) t \right]\right)$ there appear three kinds of corrections when the field operator is calculated by using the minimum-length deformed quantum mechanics. Starting from the standard (not modified at the classical level) Lagrangian, after the field quantization we get a modified dispersion relation, and besides that we find that the particle's wave function contains a small fractions of an antiparticle wave function and the backscattered wave. The result leads to interesting implications for black hole physics.

\end{abstract}




\maketitle

\section{ Introduction  }
\label{Introduction}

Minimum-length deformed quantum mechanics stems from the generalized uncertainty relation

\begin{equation}  \delta Q \delta P  \geq  \frac{1}{2} \left( 1 + \beta \delta P^2 \right) ~, \label{gur}\end{equation} proposed originally in the context of perturbative string theory as a consequence of the fact that strings cannot probe distances below the string scale (string length) \cite{String}. (We will assume natural system of units $\hbar = c = 1$ throughout this paper.) This relation immediately imposes a lower bound on position uncertainty $\delta Q_{min} = \sqrt{\beta}$ which may already imply the discreteness of the space at a fundamental level \cite{Bang:2006va}. Furthermore, Eq.\eqref{gur} was discussed by combining the basic principles of quantum theory and general relativity in the framework of {\tt Gedankenexperiments} \cite{heuristic}. Thus, the parameter $\beta$ is set by the quantum gravity scale (either by the string length or the Planck length which are of the same order) $\sqrt{\beta} \sim l_P \simeq 10^{-33}$\,cm. The minimum-length deformed quantum mechanics 

\begin{equation} [Q,\,P] = i (1 \,+\, \beta P^2)~, \label{mlqmonedim}\end{equation} underlies the generalized uncertainty relation \eqref{gur}. 
To linear order in $\beta$ this commutation relation can be solved in terms of 
standard $q,\, p$ operators $\left[ q,\, p \right] = i$
\begin{equation} Q = q~,~~~~ P = p \left[ 1 + \frac{\beta}{3}\, p^2 \right] ~.\label{onedimsoln}\end{equation}
The multidimensional generalization of Eq.\eqref{mlqmonedim} 
has the form \cite{KMM} 

\begin{equation} [Q_i,\,P_j]=i \left(\delta_{ij}+\beta P^2\delta_{ij} +
\beta^\prime P_i P_j \right)~,\label{minlengthqm}\end{equation} where both parameters $\beta$ and $\beta'$ are understood to be of the same order. For practical use it is a very useful observation that in the particular case
$\beta^\prime=2\beta$ the Eq.\eqref{minlengthqm} can be solved to linear order in $\beta$ in terms of the standard $q,\, p$ operators $\left[ q_i,\, p_j \right] = i \delta_{i j}$ \cite{Brau:1999uv}\,\footnote{For further motivation of taking $\beta^\prime=2\beta$ let us notice that the multidimensional generalization of Eq.\eqref{mlqmonedim} that preserves translation and rotation invariance and introduces a finite minimum position uncertainty in all three position variables contains only the $\beta$ parameter and leads to Eq.\eqref{brauapproach} in the leading order \cite{Kempf:1996fz, Kempf:2000ac}. }

\begin{equation} Q_i = q_i~,~~~~ P_i = p_i \left[ 1 + {\beta} \left(\mathbf{p}\right)^2 \right] ~.\label{brauapproach}\end{equation} In the 
classical limit Eq.(4) results in modified dispersion relation  
\[ \varepsilon^2 \,=\, \mathbf{p}^2 \,+\, 2\beta \mathbf{p}^4 \,+\, m^2~, \] 
which admits a simple physical interpretation: due to quantum-gravitational
fluctuations of the background metric, the energy acquires the increment $2\beta \mathbf{p}^4$. The modified algebra in Eq.\eqref{minlengthqm} has seen extensive applications in minimum length physics and gravitational 
physics. See, Refs.~\cite{Das:2008kaa,Silagadze:2009vu,Das:2009hs,Ali:2009zq,Myung:2009us} for some recent papers.



Under the influence of this modified dispersion relation the free particle wave function gets modified as \begin{eqnarray} && \exp\left(i\left[\mathbf{p}\mathbf{x} - \sqrt{{\mathbf p}^2 + m^2}\, t \right]\right) ~~\rightarrow  \nonumber  \\ && \exp\left(i\left[\mathbf{p}\mathbf{x} - \left(\sqrt{{\mathbf p}^2 + m^2} + \frac{{\beta} {\mathbf p}^4}{\sqrt{{\mathbf p}^2 + m^2}}\right) t \right]\right) ~. \label{dispwave}\end{eqnarray} Namely, the modified action as suggested in \cite{Kempf, Kempf:1996nk}

\begin{equation}\label{scaction}  \mathcal{A}[\varPhi] = - \int d^4x \, \frac{1}{2} \left[\varPhi\partial_t^2\varPhi  + \varPhi{\mathbf P}^2\varPhi 
+ m^2\varPhi^2 \right]~,\end{equation} results in the equation of motion 

\[ \partial_t^2\varPhi -
\bigtriangleup\varPhi +  2\beta \bigtriangleup\bigtriangleup\varPhi + m^2\varPhi = 0 ~,
\] which is satisfied by Eq.\eqref{dispwave}. The second related departure from the usual quantum field theory takes place when we are quantizing the field (second quantization). We will focus on this issue in what follows.  

The study of modifications to quantum field theory is motivated by a number of issues. First the generalized uncertainty principle potentially captures some features of a theory of quantum gravity and may therefore provide some insight into how quantum field theory is modified by gravity. The canonical quantization of a field theory in flat spacetime or its generalization to curved spacetimes proceeds using the harmonic oscillator for the modes. How does the resulting theory depend on this assumption, and how do important issues like the zero point energy of the modes, which are the source of the cosmological constant problem, become affected by deviations from the harmonic oscillator? In the particular modification to canonical quantization we consider (the minimum-length deformation) we find that an infrared scale necessarily enters, and has an important role in regulating the new physical effects.

\section{Preparing the setup: Free field quantization }

Let us consider a neutral scalar field  $\varPhi$ in a finite volume $l^3$

\[ H = \int\limits_{l^3} d^3x \, \frac{1}{2} \left[ \varPi^2 + \partial_{\mathbf{x}} \varPhi\partial_{\mathbf{x}} \varPhi + m^2  \varPhi^2 \right]~, \] where $\varPi = \dot{\varPhi}$. After using the Fourier expansion for $\varPi$ and $\varPhi$ \[ \varPhi(\mathbf{x}) = \frac{1}{l^3} \sum\limits_{\mathbf{p}_n} \varphi (\mathbf{p}_n)\, e^{i\mathbf{p}_n\mathbf{x}}~,~~~~  \varPi(\mathbf{x}) = \frac{1}{l^3} \sum\limits_{\mathbf{p}_n} \pi (\mathbf{p}_n)\, e^{i\mathbf{p}_n\mathbf{x}}~, \]the Hamiltonian takes the form

\[ H = \frac{1}{2 l^3} \sum\limits_{\mathbf{p}_n} \left[ \pi(\mathbf{p}_n)\pi^+(\mathbf{p}_n) + (\mathbf{p}_n^2 + m^2)\varphi(\mathbf{p}_n)\varphi^+(\mathbf{p}_n) \right] ~. \] The quantization conditions 
\begin{eqnarray} && \left[\varPhi(\mathbf{x}),\, \varPi(\mathbf{y}) \right] = i\delta (\mathbf{x} - \mathbf{y})~,~~\left[\varPhi(\mathbf{x}),\, \varPhi(\mathbf{y}) \right] =0 ~,  \nonumber \\ && \left[\varPi(\mathbf{x}),\, \varPi(\mathbf{y}) \right] = 0~,  \nonumber\end{eqnarray} for the Fourier amplitudes imply

\begin{eqnarray} && \left[\varphi(\mathbf{p}_n),\, \pi(\mathbf{p}_m)\right] = i\,l^3\,\delta_{-\mathbf{p}_n\mathbf{p}_m}~,~~\left[\varphi(\mathbf{p}_n),\, \varphi(\mathbf{p}_m)\right] = 0 ~,\nonumber \\ && \left[\pi(\mathbf{p}_n),\, \pi(\mathbf{p}_m)\right] = 0~. \nonumber \end{eqnarray} Defining 

\begin{eqnarray}&& a(\mathbf{p}_n) = \frac{1}{\sqrt{2\varepsilon_{\mathbf{p}_n}}} \left[\varepsilon_{\mathbf{p}_n}\varphi(\mathbf{p}_n) + i  \pi(\mathbf{p}_n) \right]~, \nonumber \\ && a^+(\mathbf{p}_n) = \frac{1}{\sqrt{2\varepsilon_{\mathbf{p}_n}}} \left[\varepsilon_{\mathbf{p}_n}\varphi(-\mathbf{p}_n) - i  \pi(-\mathbf{p}_n) \right]~,\nonumber \end{eqnarray} where $\varepsilon_{\mathbf{p}_n} = \sqrt{\mathbf{p}_n^2 + m^2}$, one finds 

\begin{eqnarray} && \left[a(\mathbf{p}_n),\,a^+(\mathbf{p}_m) \right] = l^3\delta_{\mathbf{p}_n\mathbf{p}_m}~,~~ \left[a(\mathbf{p}_n),\,a(\mathbf{p}_m) \right] =0~, \nonumber \\ && \left[a^+(\mathbf{p}_n),\,a^+(\mathbf{p}_m) \right] =0~. \nonumber \label{standardcreanni}\end{eqnarray} Therefore the field and momentum operators take the form 

\begin{eqnarray} \varPhi(\mathbf{x}) &=& \frac{1}{l^3} \sum\limits_{\mathbf{p}_n} \frac{1}{\sqrt{2\varepsilon_{\mathbf{p}_n}}} \left[ a(\mathbf{p}_n)e^{i\mathbf{p}_n\mathbf{x}} + a^+(\mathbf{p}_n) e^{-i\mathbf{p}_n\mathbf{x}} \right]  \nonumber ~, \nonumber \\ 
\varPi(\mathbf{x}) &=&  \frac{i}{l^3} \sum\limits_{\mathbf{p}_n}   \sqrt{\frac{\varepsilon_{\mathbf{p}_n}}{2}}  \left[   a^+(\mathbf{p}_n)e^{-i\mathbf{p}_n\mathbf{x}} -  a(\mathbf{p}_n)e^{i\mathbf{p}_n\mathbf{x}} \right]  \nonumber ~, \end{eqnarray} and the Hamiltonian reduces to  

\begin{equation} H = \frac{1}{2l^3}   \sum\limits_{\mathbf{p}_n}  \varepsilon_{\mathbf{p}_n} \left[ a^+(\mathbf{p}_n)a(\mathbf{p}_n) + a(\mathbf{p}_n)a^+(\mathbf{p}_n) \right] ~. \label{hamintermsaadagger} \nonumber\end{equation} Introducing real variables 

\begin{eqnarray} && Q_{\mathbf{p}_n} =\, \sqrt{\frac{l_{\star}}{2l^3\varepsilon_{\mathbf{p}_n} }} \left[ a(\mathbf{p}_n) + a^+(\mathbf{p}_n)\right]~,\nonumber \\ && P_{\mathbf{p}_n} = \, i\sqrt{\frac{\varepsilon_{\mathbf{p}_n}}{2l^3 l_{\star}}} \left[a^+(\mathbf{p}_n) - a(\mathbf{p}_n)\right] ~,\nonumber \end{eqnarray} the Hamiltonian splits into a sum of independent one-dimensional oscillators

\begin{equation} H =   \sum\limits_{\mathbf{p}_n} \left( \frac{l_{\star} P_{\mathbf{p}_n}^2}{2} + \frac{\varepsilon_{\mathbf{p}_n}^2 Q_{\mathbf{p}_n}^2}{2l_{\star}} \right)~.\label{oscillsum}\end{equation} In Eq.(\ref{oscillsum}) we have defined the $P_{\mathbf{p}_n}, Q_{\mathbf{p}_n}$ operators in such a way that each oscillator has a mass $l_{\star}^{-1}$. Why do we need to do so? The point is that, while in the standard quantization the energy spectrum of harmonic oscillator does not depend on its mass, for minimum-length deformed quantization the energy correction becomes mass dependent \cite{Kempf:1996fz, mloscillator}. One can conclude that the quantization of the field, suitably altered to respect the effects of a minimal length, necessarily involves some characteristic length (energy) scale $l_{\star}$ in the vein of an effective QFT. A massless free field enclosed in a box at a zero temperature has the only length scale defined by the box size. So, in this particular case one naturally infers that $l_{\star}$ should be set by the box size $l$. (It would be interesting to see how this sort of correction contributes to the Casimir force). In general, for the purpose of identifying the length scale $l_{\star},$ one may keep in mind that (in view of Eq.\eqref{mlqmonedim}) the deviation from the standard quantization becomes appreciable at high energies. Therefore it naturally suggests the identification of $l_{\star}^{-1}$ with the characteristic energy scale of the problem under consideration. This sort of reasoning is completely in the spirit of an effective QFT, we come to this point in more detail in section \ref{powercounting}. 

The Heisenberg equation of motion reads 

\[ \dot{a}(\mathbf{p}_n) = i \left[H, a(\mathbf{p}_n)\right] = - i \varepsilon_{\mathbf{p}_n}a(\mathbf{p}_n) ~, \] which can be solved as 

\[a(t,\,\mathbf{p}_n) = a(t = 0,\,\mathbf{p}_n) e^{-i\varepsilon_{\mathbf{p}_n} t} ~.\] The field and momentum operators take the form

\begin{widetext}
\begin{eqnarray} \varPhi(t,\,\mathbf{x})  & = & \frac{1}{l^3} \sum\limits_{\mathbf{p}_n} \frac{1}{\sqrt{2\varepsilon_{\mathbf{p}_n}}} \left[ a(0,\,\mathbf{p}_n) e^{i(\mathbf{p}_n\mathbf{x} - \varepsilon_{\mathbf{p}_n} t)} + a^+(0,\,\mathbf{p}_n) e^{-i(\mathbf{p}_n\mathbf{x} - \varepsilon_{\mathbf{p}_n} t)} \right]  ~, \nonumber\\
\varPi(t,\,\mathbf{x}) &=&  \frac{i}{l^3} \sum\limits_{\mathbf{p}_n}   \sqrt{\frac{\varepsilon_{\mathbf{p}_n}}{2}}  \left[   a^+(0,\,\mathbf{p}_n) e^{-i(\mathbf{p}_n\mathbf{x} - \varepsilon_{\mathbf{p}_n} t)} -  a(0,\,\mathbf{p}_n) e^{i(\mathbf{p}_n\mathbf{x} - \varepsilon_{\mathbf{p}_n} t)} \right]  \nonumber 
 ~. \end{eqnarray} 

\end{widetext} Then, we write $a(\mathbf{p}_n)$ for $a(0,\,\mathbf{p}_n)$, and similarly $a^+(\mathbf{p}_n)$ for $a^+(0,\,\mathbf{p}_n)$ in field theory and call these quantities the annihilation and creation operators, respectively.

\section{ Corrections due to minimum length deformed quantum mechanics }

In the previous section we reviewed the free field quantization of a scalar field and established our notation.
Now let us consider the implications of the minimum length deformed quantum mechanics on the field theory. For the quantization of the field we use the one-dimensional commutator in Eq.~\eqref{mlqmonedim}. This may seem a peculiar thing
to do since the field oscillators are always assumed to be harmonic oscillators. This is entirely appropriate since we want to understand the field theory excitations as particles.
Even when field theory is considered in a curved classical spacetime 
background the Fourier components of the field are quantized as a 
collection of harmonic oscillators. However in this situation it is well-known
that the notion of particle is ambiguous and leads to physically real 
processes such as Hawking radiation\cite{Birrell:1982ix}. The particle creation that occurs in such situations involves an infrared scale associated with the spacetime curvature. At the very least, one can say the introduction of the scale $\beta$ (which is related to the Planck scale and represents a deformation to the harmonic oscillators) is a method for exploring possible effects on the field quantization coming from quantum gravity. The success of field theory is due in part to the fact that other mass scales that arise in our field theories do not modify the harmonic oscillators of the second quantization. Nevertheless, gravity may be different, and in fact the reliance on harmonic oscillators in field theory has been criticized\cite{Zee:2003mt}. The introduction of this scale then necessitates the introduction of a new scale $l_{\star}$ that is not {\it a priori} related to any parameter of the theory but is rather defined by the energy scale of the problem under consideration. Speaking in a more quantitative way, the appearance of length scale $l_{\star}$ besides $\beta$ lends the possibility for introducing of a dimensionless parameter $\beta /l_{\star}^2$ that measures the deviation from the standard picture in accordance with the Eq.\eqref{mlqmonedim}.     
For each oscillator now we have 
\begin{equation} \left[Q_{\mathbf{p}_n},\,P_{\mathbf{p}_m}\right] = i\delta_{\mathbf{p}_n\mathbf{p}_m} \left(1 + \beta P^2_{\mathbf{p}_n}\right)~.\label{mlqm}\end{equation} Using the solution in Eq.\eqref{onedimsoln}
the Hamiltonian 

\[ H \,=\, \frac{l_{\star} P^2}{2} \,+\, \frac{\varepsilon^2 Q^2}{2 l_{\star}} ~,\] to the first order in $\beta$ takes the form 

\begin{eqnarray}  H  &=& \frac{l_{\star} p^2}{2} \,+\, \frac{\varepsilon^2 q^2}{2l_{\star}} \,+\, \frac{l_{\star} \beta}{3 }\, p^4  \nonumber\\ &=& \varepsilon \left( b^+b + \frac{1}{2} \right) \,+\, \frac{\beta \varepsilon^2}{12\, l_{\star}} \,(b^+ - b)^4 ~, \nonumber \end{eqnarray}
where \[ b = \sqrt{\frac{l_{\star}} {2\varepsilon}} \left[ \frac{\varepsilon}{ l_{\star} } \,q + i p \right] ~,~~~~ b^+ = \sqrt{\frac{l_{\star}} {2\varepsilon}} \left[ \frac{\varepsilon}{ l_{\star} } \, q - i p \right] ~.\] Using this Hamiltonian, from the Heisenberg equation $\dot{b} = i \left[H,\,b\right]$ one finds 

\begin{equation} \dot{b} \,=\, -i\varepsilon b \,-\, i\frac{\beta \varepsilon^2}{3 l_{\star}}\, (b^+ - b)^3 ~.\label{heqlin}\end{equation} Writing the operator $b$ to the first order in $\beta$ in the form 

\[ b = f + \beta g~,\] then Eq.\eqref{heqlin} takes the form 

\begin{equation} \dot{f} + \beta \dot{g} = -i\varepsilon (f + \beta g) \,-\, i\frac{\beta  \varepsilon^2}{3 l_{\star}}\, (f^+ - f)^3~.\label{heqfirst}\end{equation} Equating the coefficients of like powers of $\beta$ from Eq.\eqref{heqfirst} one finds

\begin{equation} \dot{f} = - i\varepsilon f~,~~ \dot{g} = -i\varepsilon g \,-\, i\frac{\varepsilon^2}{3 l_{\star}}\, (f^+ - f)^3 ~,\nonumber\end{equation} which admits the following analytic solution 

\begin{eqnarray} && f(t) = f(0) e^{-i\varepsilon t}~, \nonumber \\ && \dot{g}  = -i\varepsilon g \,-\, i\frac{\varepsilon^2}{3 l_{\star}}\, \left[ f^+(0) e^{i\varepsilon t} - f(0) e^{-i\varepsilon t}\right]^3 ~, \nonumber \\ \label{heisenbergssol} && g(t) =  \\ && e^{-i\varepsilon t } \left[ g(0) - i\frac{ \varepsilon^2}{3 \, l_{\star}} \int\limits_0^t d\tau\, e^{i\varepsilon \tau} \left\{ f^+(0) e^{i\varepsilon \tau} - f(0) e^{-i\varepsilon \tau}\right\}^3 \right]~. \nonumber \end{eqnarray} Using Eq.\eqref{heisenbergssol} to the first order in $\beta$ one can write 

\begin{eqnarray} b(t) \,=\, b(0)e^{-i\varepsilon t} \,\,- ~~~~~~~~~~~~~~~~~~~~~~~~~~~~~~~~~~~~\nonumber \\  i\frac{\beta  \varepsilon^2}{3 \, l_{\star}}\,e^{-i\varepsilon t} \int\limits_0^t d\tau\, e^{i\varepsilon \tau} \left\{ b^+(0) e^{i\varepsilon \tau} - b(0) e^{-i\varepsilon \tau}\right\}^3 ~.\nonumber\end{eqnarray} Thus, the corrected field operator takes the form 

\begin{widetext}

\begin{eqnarray} \varPhi(t,\,\mathbf{x}) \, =\, \frac{1}{l^3} \sum\limits_{\mathbf{p}_n} \frac{1}{\sqrt{2\varepsilon_{\mathbf{p}_n}}} \left[ \left(b(\mathbf{p}_n) \,-\, i\frac{\beta \varepsilon_{\mathbf{p}_n} ^2}{3 \, l_{\star}}\, \int\limits_0^t d\tau\, e^{i\varepsilon_{\mathbf{p}_n} \tau} \left[ b^+(\mathbf{p}_n) e^{i\varepsilon_{\mathbf{p}_n} \tau} - b(\mathbf{p}_n) e^{-i\varepsilon_{\mathbf{p}_n} \tau}\right]^3 \right)e^{i(\mathbf{p}_n\mathbf{x} - \varepsilon_{\mathbf{p}_n} t)} \right.\nonumber\\ \left. + \left(b^+(\mathbf{p}_n) \,+\, i\frac{\beta  \varepsilon_{\mathbf{p}_n} ^2}{3 \, l_{\star}}\, \int\limits_0^t d\tau\, e^{-i\varepsilon_{\mathbf{p}_n} \tau} \left[ b(\mathbf{p}_n) e^{-i\varepsilon_{\mathbf{p}_n} \tau} - b^+(\mathbf{p}_n) e^{i\varepsilon_{\mathbf{p}_n} \tau}\right]^3 \right)e^{-i(\mathbf{p}_n\mathbf{x} - \varepsilon_{\mathbf{p}_n} t)} \right]  \nonumber ~. \end{eqnarray}

\end{widetext} The terms from \[ \left[ b^+(\mathbf{p}_n) e^{i\varepsilon_{\mathbf{p}_n} \tau} - b(\mathbf{p}_n) e^{-i\varepsilon_{\mathbf{p}_n} \tau}\right]^3 \] that affect the matrix element $\langle 0 |\varPhi(t,\,\mathbf{x}) |\mathbf{p}_i \rangle$ are \[ e^{-i\varepsilon_{\mathbf{p}_n} \tau} b(\mathbf{p}_n)b^+(\mathbf{p}_n) b(\mathbf{p}_n) + e^{-i\varepsilon_{\mathbf{p}_n} \tau} b^2(\mathbf{p}_n)b^+(\mathbf{p}_n) ~.\] Similarly from \[ \left[ b(\mathbf{p}_n) e^{-i\varepsilon_{\mathbf{p}_n} \tau} - b^+(\mathbf{p}_n) e^{i\varepsilon_{\mathbf{p}_n} \tau}\right]^3 \] the matrix element is affected by the terms  

\[ - e^{-i\varepsilon_{\mathbf{p}_n} \tau} b(\mathbf{p}_n)b^+(\mathbf{p}_n) b(\mathbf{p}_n) - e^{-i\varepsilon_{\mathbf{p}_n} \tau} b^2(\mathbf{p}_n)b^+(\mathbf{p}_n) ~.\] Using the normalization $ a^+| n\rangle = \sqrt{n + 1} \,| n + 1 \rangle \,, ~~ a| n\rangle = \sqrt{n} \,| n - 1 \rangle$ we may write 

\begin{eqnarray}&& b^+(\mathbf{p}_n)  |\mathbf{p}_i \rangle \equiv  b^+(\mathbf{p}_n) b^+(\mathbf{p}_i) |0 \rangle  = \sqrt{1 + \delta_{ni}}\,\, |\mathbf{p}_n\,,\,\mathbf{p}_i \rangle \nonumber \\&& \mbox{and} ~~~ b^2(\mathbf{p}_n) b^+(\mathbf{p}_n)  |\mathbf{p}_i \rangle  = 2\delta_{ni} |0 \rangle~. \nonumber\end{eqnarray} Hence we find \begin{eqnarray} \langle 0 | \left[ b^+(\mathbf{p}_n) e^{i\varepsilon_{\mathbf{p}_n} \tau} - b(\mathbf{p}_n) e^{-i\varepsilon_{\mathbf{p}_n} \tau}\right]^3 |\mathbf{p}_i \rangle  &=&  3\, \delta_{ni} \, e^{-i\varepsilon_{\mathbf{p}_n} \tau}~, \nonumber\\
 \langle 0 | \left[ b(\mathbf{p}_n) e^{-i\varepsilon_{\mathbf{p}_n} \tau} - b^+(\mathbf{p}_n) e^{i\varepsilon_{\mathbf{p}_n} \tau}\right]^3  |\mathbf{p}_i \rangle  &=& - 3 \,\delta_{ni} \, e^{-i\varepsilon_{\mathbf{p}_n} \tau}~.\nonumber\end{eqnarray} The final result looks like

\begin{eqnarray}&& \langle 0 |\varPhi(t,\,\mathbf{x}) |\mathbf{p}_i \rangle  \, \propto \, e^{i(\mathbf{p}_i\mathbf{x} - \varepsilon_{\mathbf{p}_i} t)} \left(1 \,-\, i \frac{\beta  \varepsilon_{\mathbf{p}_i} ^2}{ l_{\star}}\,t\right) + \nonumber \\  && \, \frac{\beta \varepsilon_{\mathbf{p}_i}}{2l_{\star}} \, e^{-i(\mathbf{p}_i\mathbf{x} + \varepsilon_{\mathbf{p}_i} t)} \,-\, \frac{\beta \varepsilon_{\mathbf{p}_i}}{2l_{\star}}   \, e^{-i(\mathbf{p}_i\mathbf{x} - \varepsilon_{\mathbf{p}_i} t)} ~.\label{corrwavew}\end{eqnarray} From the derivation of this result it is obvious that the validity conditions for it simply imply the smallness of the corrections. For validity conditions from Eq.\eqref{dispwave} we get $\beta \mathbf{p}^4 / \varepsilon^2 \ll 1$, and from Eq.\eqref{corrwavew} $\beta  \varepsilon^2 t/l_{\star} \ll 1\,,~ \beta \varepsilon /l_{\star} \ll 1 $. We should also take into account that the generalized uncertainty relation in its minimal form Eqs.(\ref{gur},\,\ref{mlqmonedim}) as well as Brau's approach we are using, Eq.\eqref{brauapproach}, imply that $\beta p^2 \ll 1 $.

\section{Comparing with the power counting approach}
\label{powercounting}

In this section we compare developments with the effective field theory 
approach to general relativity to shed light on the physical scale $l_{\star}$.
Notice that, Eq.\eqref{mlqmonedim} can be solved exactly in terms of the standard $q,\,p$ operators \cite{Kempf:1996nk}  

\begin{equation}\label{onedimexactmodops} Q = q,~~P = \beta^{-1/2}\tan\left(p\sqrt{\beta}\right)~, \end{equation} or expanding the momentum operator in Eq.\eqref{onedimexactmodops} into series in powers of $\beta$  

\begin{equation}\label{kempfmangano} P = p + \frac{\beta}{3}\,p^3 + \frac{2\beta^2 }{15}\,p^5 + \frac{17\beta^3}{315}\,p^7 + \frac{62\beta^4}{2835}\,p^9 + O\left(\beta^5\right) ~. \end{equation} Odd powers of $p$ appear because the right hand side of Eq.\eqref{mlqmonedim} involves only integral powers of $P^2$, and consequently only even powers of $p$ will appear in the expansion of $P^2$. Introducing again the real variables 

\begin{eqnarray}&&  q_{\mathbf{k}} =\, \sqrt{\frac{l_{\star}^{-2}}{  2 \omega_{\mathbf{k}} }} \left[ b(\mathbf{k}) + b^+(\mathbf{k})\right]~,\nonumber \\&& p_{\mathbf{k}} = \, i  \sqrt{\frac{ \omega_{\mathbf{k}}l_{\star}^{-4}}{2}} \left[b^+(\mathbf{k}) - b(\mathbf{k})\right] ~,\nonumber \end{eqnarray} (for notational convenience we use in this section $(\omega_{\mathbf{k}},\,\mathbf{k})$ instead of $(\varepsilon_{\mathbf{p}},\, \mathbf{p})$ ), the Hamiltonian of a free field

\begin{eqnarray}\label{desctherulseforsmlq} H &=& \frac{1}{2}  \int d^3 k \, \omega_{\mathbf{k}}  \left[ b^+(\mathbf{k})b(\mathbf{k}) + b(\mathbf{k})b^+(\mathbf{k}) \right]   ~, \end{eqnarray} splits into a sum of independent one-dimensional oscillators 

\begin{equation} H =   \int \frac{d^3k}{l_{\star}^{-3}} \, \left( \frac{ p_{\mathbf{k}}^2}{2l_{\star}^{-1}} \,+\, \frac{l_{\star}^{-1}\omega_{\mathbf{k}}^2 q_{\mathbf{k}}^2}{2} \right)~,\label{modhambm}\end{equation} each having the mass $l_{\star}^{-1}$. Now assuming $p_{\mathbf{k}},\, q_{\mathbf{k}}$ are deformed with respect to Eqs.(\ref{onedimexactmodops}, \ref{kempfmangano}), that is, we replace $p_{\mathbf{k}} \rightarrow P_{\mathbf{k}},\, q_{\mathbf{k}} \rightarrow Q_{\mathbf{k}}$, the Hamiltonian \eqref{desctherulseforsmlq} gets modified to 

\begin{widetext}
\begin{eqnarray} && H =   \int \frac{d^3k}{l_{\star}^{-3}} \, \left( \frac{ P_{\mathbf{k}}^2}{2l_{\star}^{-1}} \,+\, \frac{l_{\star}^{-1}\omega_{\mathbf{k}}^2 Q_{\mathbf{k}}^2}{2} \right) \,=\, \nonumber \\&& \int \frac{d^3k}{l_{\star}^{-3}} \, \left( \frac{p_{\mathbf{k}}^2}{2l_{\star}^{-1}} \,+\, \frac{l_{\star}^{-1}\omega_{\mathbf{k}}^2 q_{\mathbf{k}}^2}{2} \,+\,  \beta \frac{p_{\mathbf{k}}^4}{3 \, l_{\star}^{-1}} \,+\, \beta^2 \frac{17 \, p_{\mathbf{k}}^6 }{90 \, l_{\star}^{-1}} \,+\, \beta^3 \frac{31 \, p_{\mathbf{k}}^8 }{315 \, l_{\star}^{-1}} \,+\, \beta^4 \frac{691 \, p_{\mathbf{k}}^{10} }{14175 \,l_{\star}^{-1}} \,+\, O\left(\beta^5\right) \right) \,=\, \nonumber \\ && \frac{1}{2}  \int d^3 k \, \left(\omega_{\mathbf{k}}  \left[ b^+(\mathbf{k})b(\mathbf{k}) + b(\mathbf{k})b^+(\mathbf{k}) \right] \,+\,  \frac{ \omega_{\mathbf{k}}^2 \beta}{6 \,l_{\star}^4} \left[b^+(\mathbf{k}) - b(\mathbf{k})\right]^4 \,-\,  \frac{17 \, \omega_{\mathbf{k}}^3 \beta^2}{360 \,l_{\star}^8} \left[b^+(\mathbf{k}) - b(\mathbf{k})\right]^6 \,+\, \right. \nonumber \\ && ~~~~~~~~~~~~~~~~~ \left. \frac{31\, \omega_{\mathbf{k}}^4 \beta^3}{2520\,l_{\star}^{12}} \left[b^+(\mathbf{k}) - b(\mathbf{k})\right]^8  \,-\, \frac{691\,\omega_{\mathbf{k}}^5 \beta^4}{ 226800 \,l_{\star}^{16}} \left[b^+(\mathbf{k}) - b(\mathbf{k})\right]^{10} \,+\, O\left(\beta^5\right) \right)   ~.\label{mlsqhambermanmaz} \end{eqnarray} Thus, at the second quantization level the modification amounts to the replacement of Hamiltonian \eqref{desctherulseforsmlq} with Eq.\eqref{mlsqhambermanmaz}. The perturbation Hamiltonian  

 \begin{eqnarray}&& \mathcal{H} \, = \, \frac{1}{2}  \int d^3 k \,\left[ \frac{ \omega_{\mathbf{k}}^2 \beta}{6 \,l_{\star}^4} \left[b^+(\mathbf{k}) - b(\mathbf{k})\right]^4 \,-\,  \frac{17 \, \omega_{\mathbf{k}}^3 \beta^2}{360 \,l_{\star}^8} \left[b^+(\mathbf{k}) - b(\mathbf{k})\right]^6 \,+\, \right. \nonumber \\&&~~~~~~~~~~~~~~ \left. \frac{31\, \omega_{\mathbf{k}}^4 \beta^3}{2520\,l_{\star}^{12}} \left[b^+(\mathbf{k}) - b(\mathbf{k})\right]^8  \,-\, \frac{691\,\omega_{\mathbf{k}}^5 \beta^4}{ 226800 \,l_{\star}^{16}} \left[b^+(\mathbf{k}) - b(\mathbf{k})\right]^{10} \,+\, O\left(\beta^5\right) \right]~, \end{eqnarray} can be readily written in interaction representation 
 
  \begin{eqnarray}\label{interham}&& \mathcal{H}^{Int}(t) \, = \, \frac{1}{2}  \int d^3 k \,\left[ \frac{ \omega_{\mathbf{k}}^2 \beta}{6 \,l_{\star}^4} \left[e^{i\omega_{\mathbf{k}}t}b^+(\mathbf{k}) - e^{-i\omega_{\mathbf{k}}t}b(\mathbf{k})\right]^4 \,-\,  \frac{17 \, \omega_{\mathbf{k}}^3 \beta^2}{360 \,l_{\star}^8} \left[e^{i\omega_{\mathbf{k}}t}b^+(\mathbf{k}) - e^{-i\omega_{\mathbf{k}}t}b(\mathbf{k})\right]^6 \,+\, \right. \nonumber \\&&~~~~~~~~~~~~~~ \left. \frac{31\, \omega_{\mathbf{k}}^4 \beta^3}{2520\,l_{\star}^{12}} \left[e^{i\omega_{\mathbf{k}}t}b^+(\mathbf{k}) - e^{-i\omega_{\mathbf{k}}t}b(\mathbf{k})\right]^8  \,-\, \frac{691\,\omega_{\mathbf{k}}^5 \beta^4}{ 226800 \,l_{\star}^{16}} \left[e^{i\omega_{\mathbf{k}}t}b^+(\mathbf{k}) - e^{-i\omega_{\mathbf{k}}t}b(\mathbf{k})\right]^{10} \,+\, O\left(\beta^5\right) \right]~, \end{eqnarray} where now $b^+(\mathbf{k}), b(\mathbf{k})$ are creation and annihilation operators in interaction representation. The $S$ matrix

\begin{eqnarray}  S &=& I \,-\,i\int\limits_{-\infty}^{\infty} \mathcal{H}^{Int}(\xi) d\xi \,+\, \frac{(-i)^2}{2!}\int\limits_{-\infty}^{\infty} \mathcal{T}\left[\mathcal{H}^{Int}(\xi_1)\mathcal{H}^{Int}(\xi_2)\right] d\xi_1d\xi_2 + \nonumber \\ && \frac{(-i)^3}{3!}\int\limits_{-\infty}^{\infty} \mathcal{T}\left[\mathcal{H}^{Int}(\xi_1)\mathcal{H}^{Int}(\xi_2)\mathcal{H}^{Int}(\xi_3)\right] d\xi_1d\xi_2d\xi_3  \,+ \nonumber \\ && \frac{(-i)^4}{4!}\int\limits_{-\infty}^{\infty} \mathcal{T}\left[\mathcal{H}^{Int}(\xi_1)\mathcal{H}^{Int}(\xi_2)\mathcal{H}^{Int}(\xi_3)\mathcal{H}^{Int}(\xi_4)\right] d\xi_1d\xi_2d\xi_3d\xi_4 \,+\,\cdots ,\end{eqnarray}\end{widetext}

\noindent can be used to estimate amplitudes for various processes in the spirit of an effective quantum field theory.

The correction term in the Hamiltonian \eqref{mlsqhambermanmaz} (or in Eq.\eqref{interham}) is controlled by the powers of parameter $\beta /l_{\star}^2 \equiv (l_{P}/l_{\star})^2$. Comparing it with the power counting result for gravitational scattering amplitudes obtained in the framework of an effective field theory approach to general relativity \cite{Burgess:2003jk} (section-III) \[ \mathcal{A}(E) \,=\, a_1\left(\frac{E}{m_P}\right)^2 \,+\, a_2\left(\frac{E}{m_P}\right)^4 \,+\, a_3\left(\frac{E}{m_P}\right)^6 + \cdots ~, \] where $a_i$ are numerical factors and $E$ denotes a characteristic energy scale of the process, one concludes that the length scale $l_{\star}$ should indeed be identified with $E^{-1}$. As before, in the effective field theory approach only even powers appear because the field theory involves an expansion of integral powers of squared
momenta \cite{Burgess:2003jk}.

\section{Applications to black hole physics}

We reemphasize here that the introduction of a scale $\beta$ in the field quantization procedure required us to introduce another scale $l_{\star}$. This scale enters into corrections via the dimensionless parameter $\beta/l_{\star}$ as seen in Eq.\eqref{corrwavew}. In light the discussion in the previous section, we interpret the length scale $l_{\star}$ to arise from  the energy scale of black hole physics (which is the temperature) and the associated length scale is the size of the event horizon. Precisely the same scale is at play when one quantizes a field in the black hole background (e.g. by using tortoise coordinates in which the metric appears Minkowskian).

Let us analyze the $\beta/l_{\star}$ corrections separately. Rewriting 

\begin{eqnarray} && e^{i(\mathbf{p}_i\mathbf{x} - \varepsilon_{\mathbf{p}_i} t)} \left(1 \,-\, i\frac{\beta  \varepsilon_{\mathbf{p}_i} ^2}{ l_{\star}}\,t\right) \approx   \nonumber \\ &&  e^{i(\mathbf{p}_i\mathbf{x} - \varepsilon_{\mathbf{p}_i} t)}  e^{-i\beta  \varepsilon_{\mathbf{p}_i}^2 t / l_{\star}} \,= \, e^{i\left(\mathbf{p}_i\mathbf{x} - \left[\varepsilon_{\mathbf{p}_i} +  \beta  \varepsilon_{\mathbf{p}_i}^2  / l_{\star}\right] t \right)}  ~, ~\label{enincr}\end{eqnarray} we see that energy gets increased leading to the modified dispersion relation \begin{equation}  \varepsilon = \sqrt{\mathbf{p}^2 + m^2} + \beta \, \frac{\mathbf{p}^2 + m^2}{l_{\star}}~.\label{moddisprel}\end{equation}

The second term (on the right-hand side) in Eq.\eqref{corrwavew} represents a reflected wave. The probability of reflection is proportional to $\propto \left|\beta \varepsilon(\mathbf{p}) / l_{\star}\right|^2$. This phenomenon provides a mechanism for obtaining the black hole remnants. It tells us that during the black hole evaporation there is a backscattered flux as well that falls into the black hole. For Hawking radiation $T \propto r_g^{-1}$, where $r_g$ denotes the radius of the black hole, that is, $T \propto m_P^2/M$, where $M$ stands for the black hole mass. So, we get $dM \propto - m_P^2 dT /T^2$. Because of backscattered flux the mass increment of the black hole takes the form $dM_{+} \propto \left|\beta \varepsilon / l_{\star}\right|^2dM$. By taking into account that $\varepsilon \propto T$ and the appropriate energy scale at hand for defining $l_{\star}$ is $T$, that is, $l_{\star} \sim  T^{-1}$, we get $dM_{+} \propto T^2 dT /m_P^2$. So we are imagining that the field quantization depends on the Planck scale through the parameter $\beta$, and then for the black hole the appropriate infrared length scale arises from the scale of spacetime curvature, i.e. the radius $r_g$ of the black hole. Thus, for the black hole mass we get \begin{equation}  M \,=\, \frac{1}{8\pi}\, \frac{m_P^2}{ T} \,+\, \delta \, \frac{T^3}{m_P^2}~,\label{bhmass} \end{equation} where $\delta$ is a numerical factor of order unity. The first term in Eq.\eqref{bhmass} is much greater in comparison with the second one as long as $T^4 \ll m_P^4$. But when the temperature approaches the Planck scale, the second term becomes of the order of the first one. The black hole mass reaches its minimum $M_{min} \sim m_P$ for $T \sim m_P$. The mechanism of stabilization is clear. Let us notice that the question of black hole remnants was addressed in a heuristic way immediately by using the generalized uncertainty relation in \cite{Adler:2001vs}. The temperature of black hole radiation obtained heuristically from the generalized uncertainty relation becomes imaginary when the black hole mass drops below the Planck scale. That was interpreted as the reason for the emission halting. We see that emission does not halt as such; instead, the backscattered flux becomes comparable to the outgoing flux when the black hole evaporates down to the Planck mass, and that serves as a mechanism of stabilization. However it should be remembered that our expectations are that applying just the modifications to the uncertainty principle are likely insufficient for describing the physics of a Planck mass black hole. In fact the validity conditions for our perturbative calculation are violated if one goes to the extreme case of such a low mass black hole. Furthermore, the expectation is that Eq.\eqref{mlqm} represents at best a truncated expression for the commutator. The full expression presumably involves a (convergent) power series in $P^2$ whose effects beyond first order in $\beta$ escape our perturbative analysis.

Let us notice that Planck mass black hole remnants may be an interesting candidate for the dark matter \cite{Chen:2002tu}. Such primordial black holes with a sufficient number for the dark matter can be produced during the inflation \cite{primbh}. 

It should also be noticed that in general the effect of a backscattering is not new for black hole emission. Because of the (usual) backscattering of the Hawking radiation off the curved background, the spectrum of the black hole radiation is not precisely of the form of a black body radiation but gets modified with the gray body factor

\[N_{\omega l m} = \frac{1}{2\pi}\, \frac{1}{ e^{4\pi \omega r_g} -1}~,~~ \Rightarrow ~~ N_{\omega l m} = \frac{1}{2\pi}\, \frac{\Gamma_{\omega l}}{ e^{4\pi \omega r_g} -1}~, \] where the gray body factor, $\Gamma_{\omega l}$, indicates the decay (supression) of the outgoing flux by a factor $1 - \Gamma_{\omega l}$ for this part of the outgoing flux becomes reflected and falls back to the black hole \cite{Birrell:1982ix}. It is important to emphasize that the appearance of gray body factor does not affect the thermal character of the radiation (or otherwise, black hole still remains a thermal object) for this gray body factor works in both directions, that is, for outgoing and incoming fluxes equivalently. One important difference between the compared backscattering effects is the dependence on energy. For the traditional effect (calculated using quantum field theory in a curved background geometry) one has $\Gamma \to 1$ for frequencies large compared to the inverse black hole radius, whereas the effect uncovered here grows with energy.

The thermodynamic interpretation of the black hole can be maintained in the presence of the order $\beta$ corrections. Using the Eq.\eqref{bhmass} and the formula $dS = dM/T$  we get the following correction to the black hole entropy 

\begin{equation} S = \pi\left(\frac{r_g}{l_P}\right)^2  + \eta \left(\frac{l_P}{r_g}\right)^2~,\label{inversearea}\end{equation} where $\eta = 3\delta/32\pi^2$. 

It is easy to see that the increment of energy given by Eq.\eqref{enincr} (or otherwise, the modified dispersion relation Eq.\eqref{moddisprel}) results in the logarithmic correction to the black hole entropy. Taking $\varepsilon \propto T\,,~l_{\star} \sim  T^{-1}$, Eq.\eqref{enincr} tells us that black hole emission temperature gets increased $T \rightarrow T + \beta T^3$. Hence, to the first order in $\beta$ the entropy $dS = dM/T \rightarrow dM/T - \beta T dM$ acquires a logarithmic correction 

\[ S = \pi\left(\frac{r_g}{l_P}\right)^2  - \gamma  \ln\left(\frac{r_g}{l_P}\right)~,\] where $\gamma$ is a (positive) number of order unity. Combining with Eq.\eqref{inversearea} we get a well known entropy expression

\[ S = \pi\left(\frac{r_g}{l_P}\right)^2  - \gamma  \ln\left(\frac{r_g}{l_P}\right) + \eta \left(\frac{l_P}{r_g}\right)^2 + \mbox{const.} ~,\] obtained in loop quantum gravity \cite{lqgbhent} and in a tunneling formalism approach to the black hole emission \cite{tunbhent}. The logarithmic corrections seem to be a generic property of black holes\cite{Das:2001ic},  but now we have clear understanding of the physics behind each term (at least as far as a heuristic explanations affords).

The third term (on the right-hand side) in Eq.\eqref{corrwavew} could be interpreted as indicating the possibility of particle transition into antiparticle \cite{Hawking:1979hw, Hawking:1979pi, Page:1980qm} in the framework of minimum-length deformed quantum mechanics. But strictly speaking, in the case of a real, fundamental scalar there is no antiparticle as such. It's worth noticing that the effect of particle-antiparticle transition is absent for charged scalar field as well, see the discusion below. So, this effect makes sense only if $ \varPhi$ is taken as an effective field describing a composite neutral particle. The probability of this process is equal to the likelihood of the backscattering effect $\propto \left|\beta \varepsilon(\mathbf{p}) / l_{\star}\right|^2$. The presence of antiparticle content in the plane wave might also suggest the presence of particle-antiparticle creation.

One may wonder about the validity of the above discussion for charged particles. The generalization of the above discussion to this case is straightforward. For a complex scalar field the field operator and the Hamiltonian take the form \cite{LL}

\[ \varPhi(\mathbf{x}) = \frac{1}{l^3} \sum\limits_{\mathbf{p}_n} \frac{1}{\sqrt{2\varepsilon_{\mathbf{p}_n}}} \left[ c(\mathbf{p}_n)e^{i\mathbf{p}_n\mathbf{x}} + d^+(\mathbf{p}_n) e^{-i\mathbf{p}_n\mathbf{x}} \right]  ~ , \]

\begin{equation} H = \frac{1}{2l^3}   \sum\limits_{\mathbf{k}_n}  \varepsilon_{\mathbf{p}_n} \left[ c^+(\mathbf{p}_n)c(\mathbf{p}_n) + d(\mathbf{p}_n)d^+(\mathbf{p}_n) \right] ~, \nonumber \end{equation} where the particle operators $c(\mathbf{p}),\,c^+(\mathbf{p})$ commute with antiparticle operators $d(\mathbf{p}),\,d^+(\mathbf{p})$. From now on one can immediately use the above discussion separately for $c(\mathbf{p}),\,c^+(\mathbf{p})$ and $d(\mathbf{p}),\,d^+(\mathbf{p})$ pairs respectively. Therefore, the field operator takes the form

\begin{widetext}

\begin{eqnarray} \varPhi(t,\,\mathbf{x}) \, =\, \frac{1}{l^3} \sum\limits_{\mathbf{p}_n} \frac{1}{\sqrt{2\varepsilon_{\mathbf{p}_n}}} \left[ \left(c(\mathbf{p}_n) \,-\, i\frac{\beta \varepsilon_{\mathbf{p}_n} ^2}{3 \, l_{\star}}\, \int\limits_0^t d\tau\, e^{i\varepsilon_{\mathbf{p}_n} \tau} \left[ c^+(\mathbf{p}_n) e^{i\varepsilon_{\mathbf{p}_n} \tau} - c(\mathbf{p}_n) e^{-i\varepsilon_{\mathbf{p}_n} \tau}\right]^3 \right)e^{i(\mathbf{p}_n\mathbf{x} - \varepsilon_{\mathbf{p}_n} t)} \right.\nonumber\\ \left. + \left(d^+(\mathbf{p}_n) \,+\, i\frac{\beta  \varepsilon_{\mathbf{p}_n} ^2}{3 \, l_{\star}}\, \int\limits_0^t d\tau\, e^{-i\varepsilon_{\mathbf{p}_n} \tau} \left[ d(\mathbf{p}_n) e^{-i\varepsilon_{\mathbf{p}_n} \tau} - d^+(\mathbf{p}_n) e^{i\varepsilon_{\mathbf{p}_n} \tau}\right]^3 \right)e^{-i(\mathbf{p}_n\mathbf{x} - \varepsilon_{\mathbf{p}_n} t)} \right]  \nonumber ~. \end{eqnarray}

\end{widetext} Similarly the cases of particles with spin involve straightforward generalizations. It is easy to see that as compared to Eq.\eqref{corrwavew}, in the case of a charged field there is neither a reflected component nor an antiparticle wavefunction. For stabilizing the mini black holes (to get black hole remnants) it is important to know whether the higher order corrections to the field operator in $\beta$ lead to the backscattering effect for charged particles or not. Besides that, it is important to estimate the effect of electrostatic attraction between the emitted particle and antiparticle. In general this effect causes the average emission rate and power to be lower than for otherwise similar uncharged particles \cite{Page:1977um}. In the case of Planck size black hole one has to account also for the fact that the electric coupling increases with energy and in general one may also expect the electrostatic force at this scale to be strong enough to cause the backcapture of the charged particles by the black hole. 
 
Many papers have addressed the phenomenological implications of generalized uncertainty relation for the mini black holes in the framework of extra-dimensional models with low quantum gravity scale. What the above discusion tells us is that in such a case the emission rate for the uncharged particles should be highly suppressed. 

The modifications to the plane wave solutions arising here from the generalized uncertainty principle can be compared to the more conventional studies of quantum field theory in a curved spacetime (with canonical quantization). Many of the same effects appear here such as contributions similar to backscattering off the curved geometry in Hawking radiation, the mixing of positive and negative frequency modes and particle creation. Whether these similarities can and should be identified (or whether possible effects from minimum-length considerations serve as a supplement) may be an interesting topic for future investigations. Finally it is unclear whether there is a path integral approach to quantization which captures the same physical effects of the minimum-length uncertainty relation. 

\vspace{0.06cm}
{\tt \bf  Acknowledgments:} This work was supported in part by DOE grant DE-FG02-91ER40661 and by the Indiana University Center for Spacetime Symmetries. Finally, we express our gratitude to the Referee for many useful questions and comments.


\begin{thebibliography}{10}

\bibitem{String}
G.~Veneziano,
Europhys.\ Lett.\  {\bf 2}, 199 (1986);

D.~J.~Gross and P.~F.~Mende,
Nucl.\ Phys.\  B {\bf 303}, 407 (1988);

D.~Amati, M.~Ciafaloni and G.~Veneziano,
Phys.\ Lett.\  B {\bf 216}, 41 (1989);

K.~Konishi, G.~Paffuti and P.~Provero,
Phys.\ Lett.\  B {\bf 234}, 276 (1990);

R.~Guida, K.~Konishi and P.~Provero,
Mod.\ Phys.\ Lett.\  A {\bf 6}, 1487 (1991).


\bibitem{Bang:2006va}
  J.~Y.~Bang and M.~S.~Berger,
  Phys.\ Rev.\  D {\bf 74}, 125012 (2006)
  [arXiv: gr-qc/0610056];


  J.~Y.~Bang and M.~S.~Berger,
  Phys.\ Rev.\  A {\bf 80}, 022105 (2009)
  [arXiv: 0811.0838 [quant-ph]].


\bibitem{heuristic}
M.~Maggiore,
Phys.\ Lett.\  B {\bf 304}, 65 (1993) [arXiv: hep-th/9301067];

F.~Scardigli,
Phys.\ Lett.\  B {\bf 452}, 39 (1999) [arXiv: hep-th/9904025];

R.~J.~Adler and D.~I.~Santiago,
Mod.\ Phys.\ Lett.\  A {\bf 14}, 1371 (1999) [arXiv:
gr-qc/9904026].


\bibitem{KMM}
A.~Kempf, G.~Mangano and R.~B.~Mann,
Phys.\ Rev.\  D {\bf 52}, 1108 (1995) [arXiv: hep-th/9412167].

\bibitem{Brau:1999uv}
  F.~Brau,
  J.\ Phys.\ A  {\bf 32}, 7691 (1999)
  [arXiv: quant-ph/9905033].

\bibitem{Kempf:1996fz}
  A.~Kempf,
  J.\ Phys.\ A  {\bf 30}, 2093 (1997)
  [arXiv: hep-th/9604045].


\bibitem{Kempf:2000ac}
  A.~Kempf,
  Phys.\ Rev.\  D {\bf 63}, 083514 (2001)
  [arXiv: astro-ph/0009209].





\bibitem{Das:2008kaa}
  S.~Das and E.~C.~Vagenas,
  Phys.\ Rev.\ Lett.\  {\bf 101}, 221301 (2008)
  [arXiv: 0810.5333 [hep-th]].

\bibitem{Silagadze:2009vu}
  Z.~K.~Silagadze,
  Phys.\ Lett.\  A {\bf 373}, 2643 (2009)
  [arXiv: 0901.1258 [gr-qc]].

\bibitem{Das:2009hs}
  S.~Das and E.~C.~Vagenas,
  Can.\ J.\ Phys.\  {\bf 87}, 233 (2009)
  [arXiv: 0901.1768 [hep-th]].

\bibitem{Ali:2009zq}
  A.~F.~Ali, S.~Das and E.~C.~Vagenas,
  Phys.\ Lett.\  B {\bf 678}, 497 (2009)
  [arXiv: 0906.5396 [hep-th]].

\bibitem{Myung:2009us}
  Y.~S.~Myung,
  Phys.\ Lett.\  B {\bf 684}, 158 (2010)
  [arXiv: 0908.4132 [hep-th]].

\bibitem{Kempf}

A.~Kempf,
  arXiv: hep-th/9405067;

 A.~Kempf,
  J.\ Math.\ Phys.\  {\bf 38}, 1347 (1997)
  [arXiv: hep-th/9602085];

A.~Kempf,
  Phys.\ Rev.\  D {\bf 54}, 5174 (1996)
  [Erratum-ibid.\  D {\bf 55}, 1114 (1997)]
  [arXiv: hep-th/9602119];


\bibitem{Kempf:1996nk}
  A.~Kempf and G.~Mangano,
  Phys.\ Rev.\  D {\bf 55}, 7909 (1997)
  [arXiv: hep-th/9612084].



\bibitem{mloscillator}

 
L.~N.~Chang, D.~Minic, N.~Okamura and T.~Takeuchi,
Phys.\ Rev.\  D {\bf 65}, 125027 (2002) [arXiv: hep-th/0111181].


\bibitem{Birrell:1982ix}
  N.~D.~Birrell and P.~C.~W.~Davies, "Quantum Fields In Curved Space,"
(Cambridge Univ. Press, New York, 1982).

\bibitem{Zee:2003mt}
  A.~Zee, "Quantum field theory in a nutshell",
{(Princeton, UK: Princeton University Press, 2010)}.


\bibitem{Burgess:2003jk}
  C.~P.~Burgess,
  Living Rev.\ Rel.\  {\bf 7}, 5 (2004)
  [arXiv: gr-qc/0311082].


\bibitem{Adler:2001vs}
  R.~J.~Adler, P.~Chen and D.~I.~Santiago,
  Gen.\ Rel.\ Grav.\  {\bf 33}, 2101 (2001)
  [arXiv: gr-qc/0106080].



\bibitem{Chen:2002tu}
  P.~Chen and R.~J.~Adler,
  Nucl.\ Phys.\ Proc.\ Suppl.\  {\bf 124}, 103 (2003)
  [arXiv: gr-qc/0205106];



  P.~Chen,
  New Astron.\ Rev.\  {\bf 49}, 233 (2005)
  [arXiv: astro-ph/0406514].



\bibitem{primbh}

  P.~Ivanov, P.~Naselsky and I.~Novikov,
  Phys.\ Rev.\  D {\bf 50}, 7173 (1994).


  J.~Garcia-Bellido, A.~D.~Linde and D.~Wands,
  Phys.\ Rev.\  D {\bf 54}, 6040 (1996)
  [arXiv: astro-ph/9605094].


  J.~Yokoyama,
  Astron.\ Astrophys.\  {\bf 318}, 673 (1997)
  [arXiv: astro-ph/9509027].


  M.~Kawasaki and T.~Yanagida,
  Phys.\ Rev.\  D {\bf 59}, 043512 (1999)
  [arXiv: hep-ph/9807544].


  M.~Kawasaki, T.~Takayama, M.~Yamaguchi and J.~Yokoyama,
  Phys.\ Rev.\  D {\bf 74}, 043525 (2006)
  [arXiv: hep-ph/0605271].


\bibitem{lqgbhent}


  C.~Rovelli,
  Phys.\ Rev.\ Lett.\  {\bf 77}, 3288 (1996)
  [arXiv: gr-qc/9603063];


  A.~Ashtekar, J.~Baez, A.~Corichi and K.~Krasnov,
  Phys.\ Rev.\ Lett.\  {\bf 80}, 904 (1998)
  [arXiv: gr-qc/9710007].


\bibitem{tunbhent}

  A.~J.~M.~Medved and E.~C.~Vagenas,
  Mod.\ Phys.\ Lett.\  A {\bf 20}, 1723 (2005)
  [arXiv: gr-qc/0505015];


  J.~Zhang,
  Phys.\ Lett.\  B {\bf 668}, 353 (2008)
  [arXiv: 0806.2441 [hep-th]].

\bibitem{Das:2001ic}
  S.~Das, P.~Majumdar and R.~K.~Bhaduri,
  Class.\ Quant.\ Grav.\  {\bf 19}, 2355 (2002)
  [arXiv: hep-th/0111001].


\bibitem{Hawking:1979hw}
  S.~W.~Hawking, D.~N.~Page and C.~N.~Pope,
  Phys.\ Lett.\  B {\bf 86}, 175 (1979).


\bibitem{Hawking:1979pi}
  S.~W.~Hawking, D.~N.~Page and C.~N.~Pope,
  Nucl.\ Phys.\  B {\bf 170}, 283 (1980).


\bibitem{Page:1980qm}
  D.~N.~Page,
  Phys.\ Lett.\  B {\bf 95}, 244 (1980).



\bibitem{LL}

 V.~B.~Berestetskii, E.~M.~Lifshitz and L.~P.~Pitaevskii, "Quantum electrodynamics" (Landau and Lifshitz Course of Theoretical Physics - Vol. IV; Oxford: Pergamon Press, 1982). 




\bibitem{Page:1977um}
  D.~N.~Page,
  Phys.\ Rev.\  D {\bf 16}, 2402 (1977).


\end{thebibliography}
\end{document}